\documentclass[12pt,amsmath,amssymb]{iopart}

\usepackage[cp1251]{inputenc}
\usepackage[english]{babel}
\usepackage{graphicx}
\begin{document}

\title{Peak Doubling in SPDC Coincidence Spectra with a Short-Pulse Pump}

\author{M V Fedorov$^{1\,*}$, J M Mikhailova$^{1,\,2}$ and P A Volkov$^1$}

\address{$^1$A.M.~Prokhorov General Physics Institute,
 Russian Academy of Sciences, 38 Vavilov st., Moscow, 119991, Russia\\
 $^2$ Max-Planck-Institut f{\" u}r Quantenoptik, Hans-Kopfermann-Strasse 1,
D-85748, Garching, Germany\\
$^*$E-mail: fedorovmv@gmail.com}
\begin{abstract}
We describe a double-peak structure of the coincidence spectrum of biphoton states in the process of spontaneous parametric down-conversion with a pump having the form of short pulses. The effect is shown to arise owing to the obligatory symmetry of bihoton wave functions, as wave functions describing states of two bozons obeying the Bose-Einstein statistics. Parameters of the peaks are found and conditions necessary for experimental observation of the effect are determined.
\end{abstract}

\maketitle

\section{Introduction}
As known \cite{Angular,Spectral}, biphoton states generated in Spontaneous Parametric Down-Conversion (SPDC) can be highly entangled in continuous variables of signal and idler photons, such as frequencies or angular variables. An efficient way of finding the degree of entanglement from experimental data consists in measuring single-particle and coincidence angular distributions or spectra, finding their widths (correspondingly, $\Delta\omega^{(s)}$ and $\Delta\omega^{(c)}$ for spectral distributions) and using their ratio $R=\Delta\omega^{(s)}/\Delta\omega^{(c)}$ as the entanglement quantifier. As  shown in Refs. \cite{2004, 2006}, for double-Gaussian distributions $R\equiv K$, where $K$ is the Schmidt parameter \cite{Grobe, Knight}. Both parameters $K$ and $R$ can be used as entanglement quantifiers only for pure biphoton states. The approximate equality $R\approx K$ assumes that both single-photon and coincidence distributions have  smooth one-peak forms and can be approximated by Gaussian curves. In this paper we show that there are cases when both assumptions about purity of biphoton states and about Gaussian-like shape of the coincidence distributions are invalid. The key point of this consideration is related to the obligatory symmetry of biphoton wave functions with respect to transpositions of the photon variables. Importance of this feature of any two-particle states with identical (indistinguishable) particles was discussed in details in Ref. \cite{Archive} in application to systems with discrete variables (biphoton qutrits and ququarts). Here we consider the case of continuous variables (frequencies of photons) in the SPDC processes with collinear phase-matching and with the pump having the form of a short pulse.

\subsection{Type-I collinear phase-matching}

Let us consider first the case of the type-I collinear phase-matching. In this process the pump propagates in a nonlinear crystal as an extraordinary wave, and some of its photons decay for two photons propagating in the same direction ($z$-axis) as ordinary waves,   $e\rightarrow o+o$. Both emitted photons have the same polarization, e.g., the horizontal one, whereas the pump is vertically polarized. We assume also that the pump is a Gaussian pulse of a duration $\tau$. Under these conditions the polarization-spectral wave function of two emitted photons has the form \cite{Spectral}
\begin{equation}
 \label{WF-I}
 \Psi\propto \left(1\atop 0\right)_1\left(1\atop 0\right)_2\exp\left(-\frac{(\omega_1+\omega_2-\omega_0)^2\tau^2}{8\ln
 2}\right)
  {\rm  sinc}\left\{\frac{L}{2} \Delta(\omega_1, \omega_2, \phi)
 \right\},
\end{equation}
where $\omega_1$ and $\omega_2$ are frequencies of emitted photons, $\omega_0$ is the central frequency of the pump, $\phi$ is the angle between the optical axis of a crystal and the z-axis (direction of propagation of all photons), $L$ is the length of a crystal in the propagation, $z$,  direction. Two columns in Eq. (\ref{WF-I}), $\left(1\atop 0\right)_1$ and $\left(1\atop 0\right)_2$, are polarization parts of the biphoton wave function with indices 1 and 2 referring two emitted photons.  $\Delta(\omega_1, \omega_2, \phi)$ is the phase mismatch
\begin{eqnarray}
 \nonumber
 \Delta(\omega_1, \omega_2, \phi)=k_p(\omega_1+\omega_2, \phi)-k_1(\omega_1)-k_2(\omega_2)\\
 \label{Mismatch-1}
 =n_e(\omega_1+\omega_2, \phi)\frac{\omega_1+\omega_2}{c}
 -n_o(\omega_1)\frac{\omega_1}{c}-n_o(\omega_2)\frac{\omega_2}{c},
\end{eqnarray}
where and $n_o(\omega)$ is the isotropic index of an ordinary wave (emitted photons), and $n_e(\omega, \phi)$ is the refractive index of the pump
\begin{equation}
 \label{np}
 n_e(\omega, \phi)= \frac{n_o(\omega)n_e(\omega)}{[(n_o(\omega)\sin\phi)^2 + (n_e(\omega)\cos\phi)^2]^{1/2}},
\end{equation}
$n_e(\omega)$ is the refractive index of an extraordinary wave that would propagate  along the optical axis. The functions $n_o(\omega)$ and $n_e(\omega)$ are determined by the well known Sellmeier formulas \cite{Sellmeier}.

Without any further detailing, it's clear from Eqs. (\ref{WF-I}) and (\ref{Mismatch-1}) that in the case of the type-I phase-matching the polarization-spectral biphoton wave function is symmetric with respect to the transposition of particle variables $1\rightleftharpoons 2$ and does not need any additional symmetrization. Actually, symmetry of the wave function (\ref{WF-I}) is related to the fact that both emitted photons belong to the same type of waves, ordinary waves.

Another general comment concerns the polarization degree of freedom of emitted photons. The polarization part of the wave function (\ref{WF-I}) is factorized for the product of two terms depending separately on polarization variables of photons 1 and 2. Also, it is factorized with respect to the part depending on frequency variables. For this reason, in the SPDC process with the type-I phase-matching, there is no polarization entanglement. Moreover, a pure biphoton state characterized by the wave function of Eq. (\ref{WF-I}) remains pure even if it is reduced with respect to the polarization variables. In other words, the bihoton state has only purely frequency entanglement, and for its investigation polarization part of the wave function simply can be  ignored. This contrasts with the case of the type-II phase-matching considered in the next section.
\begin{figure}[h]
\centering\includegraphics[width=8.5cm]{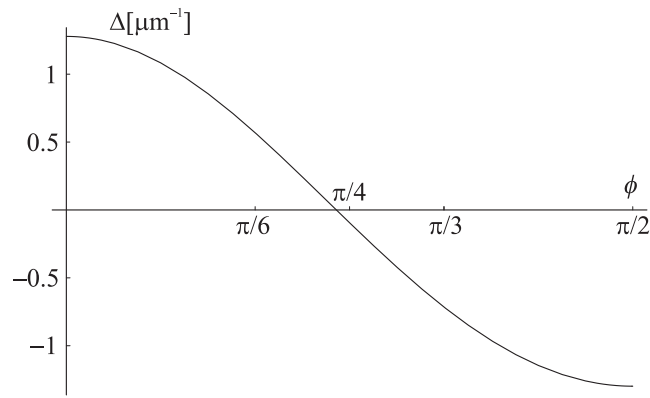}
\caption{{\protect\footnotesize {Angular dependence of the phase mismatch at central frequencies, $\Delta(\omega_0/2, \omega_0/2, \phi)$ .}}}\label{Fig1}
\end{figure}

To specify a little bit further the most typical experimentally met conditions, let us assume that the SPDC process under consideration is degenerate with respect to central frequencies of spectra of emitted photons, $\omega_1^{(0)}=\omega_2^{(0)}=\omega_0/2$. Such situation occurs if at these frequencies the phase mismatch turns zero, $\Delta(\omega_0/2, \omega_0/2, \phi)=0$, which can be provided by an appropriate choice of the angle $\phi$. Figure \ref{Fig1} shows the dependence of the phase mismatch at central frequencies on $\phi$. Here and below all calculations  are made for a crystal LiIO$_3$ and the central wavelength of the pump $\lambda_0= 0.3975\mu{\rm m}$. As can be found from Figure \ref{Fig1}, the degeneracy condition  $\Delta\left(\omega_0/2, \omega_0/2, \phi\right)=0$ is satisfied at $\phi=\phi_0=42.904^\circ$.

\begin{figure}[h]
\centering\includegraphics[width=8.5cm]{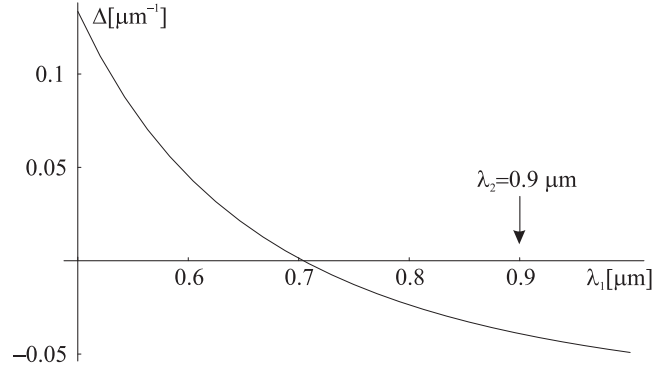}
\caption{{\protect\footnotesize {Mismatch (\ref{Mismatch-1}) as a function of $\lambda_1$ at $\lambda_2=0.9 \mu{\rm m}$.}}}\label{Fig2}
\end{figure}

\begin{figure}[h]
\centering\includegraphics[width=8.5cm]{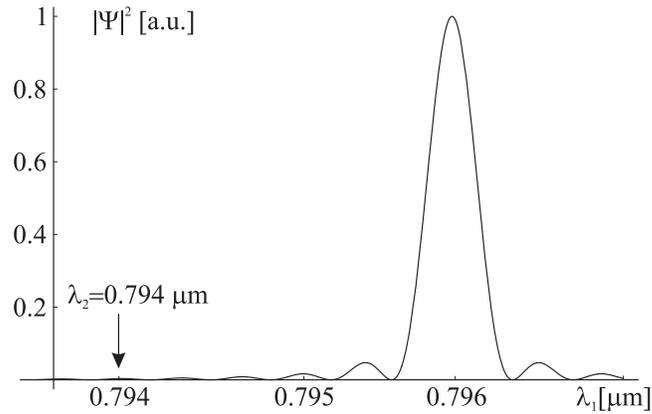}
\caption{{\protect\footnotesize {Coincidence spectrum, $|\Psi(\lambda_1,\lambda_2)|^2$ (in arbitrary units), at $\lambda_2=const =0.794 \mu{\rm m}$; the type-I collinear degenerate phase-matching, pulse duration and central wavelength of the pump $\tau=50 fs$ and $\lambda_0=0.795 \mu{\rm m}$, LiIO$_3$ crystal of the length $L=1 cm$.}}}\label{Fig3}
\end{figure}

Figure \ref{Fig2} shows a typical spectral dependence of the mismatch $\Delta$ at $\phi=\phi_0$ but frequencies different from those providing exact phase-matching. In this and all further figures all calculated spectra are plotted in dependence on wavelengths rather than frequencies, which is more appropriate for comparison with existing and possible experiments.
As seen from Figure \ref{Fig2}, the mismatch is a monotonous function of $\lambda _1$ at a given value of $\lambda _2$ (or of a frequency $\omega_1$ at a given $\omega_2$). For this reason, the mismatch  turns zero at only one value of $\lambda _1$ (or $\omega_1=2\pi c/\lambda_1$). As the zero mismatch corresponds to the maximal value of the sinc-function in Eq. (\ref{WF-I}), the coincidence spectrum has in this case a simple single-peak structure (Figure \ref{Fig3}), in contrast to the spectrum arising in the case of the type-II phase-matching discussed below.

\section{Type-II collinear phase-matching}

\begin{figure}[h]
\centering\includegraphics[width=8.5cm]{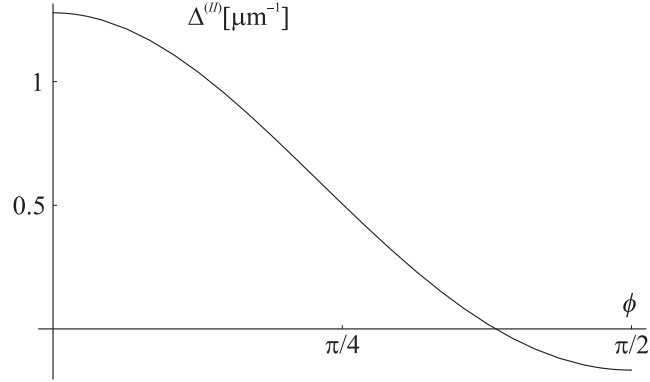}
\caption{{\protect\footnotesize {Angular dependence of the phase mismatch at central frequencies\\ $\Delta^{(II)}(\phi)$ (\ref{DEL-II-centr}).}}}\label{Fig4}
\end{figure}

In the case of the type-II phase-matching, the pump and one of two emitted photons are extraordinary whereas the second emitted photon is of the ordinary waves, i.e., the SPDC process has the form $e\rightarrow o+e$. As previously, let us assume that the process is collinear and degenerate with respect to the central frequencies of all photons. Let, as previously, polarizations of extraordinary and ordinary photons be, correspondingly, vertical and horizontal. The phase mismatch of such process at central frequencies is given by
\begin{eqnarray}
 \nonumber
 \Delta^{(II)}(\phi)=k_p(\omega_0/2, \phi)-k_o(\omega_0/2)-k_e(\omega_0/2, \phi)\\
 \label{DEL-II-centr}
 =n_e(\omega_0, \phi)\frac{\omega_0}{c}
 -n_o\left(\frac{\omega_0}{2}\right)\frac{\omega_0}{2c} -n_e\left(\frac{\omega_0}{2},\phi\right)\frac{\omega_0}{2c}.
\end{eqnarray}
For LiIO$_3$ and $\lambda_0=0.39975 \mu{\rm m}$ the function $\Delta^{(II)}(\phi)$ is plotted in Figure \ref{Fig4}.
The exact-phase-matching angle at which $\Delta^{(II)}(\phi)=0$ found from Figure \ref{Fig4} is $\phi_0^{(II)}=68.845^\circ$. Both this value and the curve of Figure \ref{Fig4} rather pronouncedly differ from $\phi_0^{(I)}=42.904^\circ$ and the curve of Figure \ref{Fig1} occurring in the case of the type-I phase-matching.

\begin{figure}[h]
\centering\includegraphics[width=8.5cm]{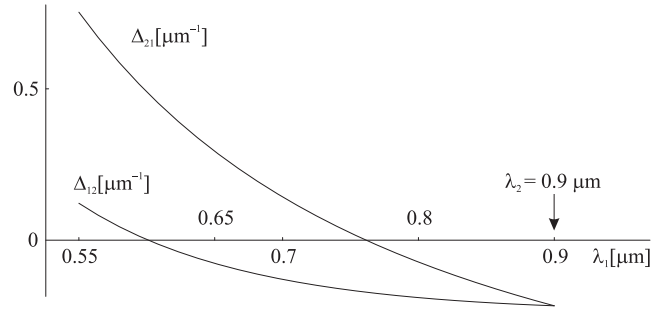}
\caption{{\protect\footnotesize {Mismatches
$\Delta_{12}$ (\ref{Del12}) and $\Delta_{21}$ (\ref{Del21})
as functions of $\lambda_1$ at $\lambda_2=0.9 \mu{\rm m}$ for LiIO$_3$ and $\lambda_0=0.795 \mu{\rm m}$
.}}}\label{Fig5}
\end{figure}

At frequencies of emitted photons $\omega_1$ and $\omega_2$ differing from $\omega_0/2$, the phase mismatch $\Delta$  of the type-II SPDC process is asymmetric with respect to transposition of photo variables (see Eqs. (\ref{Del12}) and (\ref{Del21}) below). The same is true for wave functions constructed from a given $\Delta$. But as photons are bozons, their wave functions must be symmetric \cite{AandB}. Symmetrization of wave the function is provided by summation of its expression with a given $\Delta$ and the same expression with transposed frequency and polarization variables of photon :
\begin{eqnarray}
 \nonumber
 \Psi\propto \exp\left(-\frac{(\omega_1+\omega_2-\omega_0)^2\tau^2}{8\ln 2}\right)\\
 \label{symmWF}
 \times\left\{\left(1\atop 0\right)_1\left(0\atop 1\right)_2{\rm sinc}\left[\frac{L}{2}\Delta_{12}\right]
 +\left(1\atop 0\right)_2\left(0\atop 1\right)_1{\rm sinc}\left[\frac{L}{2}\Delta_{21}\right]\right\},
\end{eqnarray}
where
\begin{eqnarray}
 \nonumber
 \Delta_{12}(\omega_1,\omega_2)=k_p\left(\omega_1+\omega_2, \phi_0^{(II)}\right)-k_o(\omega_1)-k_e\left(\omega_2, \phi_0^{(II)}\right)\\
 \label{Del12}
 =n_e\left(\omega_1+\omega_2,\,\phi_0^{(II)}\right)\frac{\omega_1+\omega_2}{c}
 -n_o(\omega_1)\frac{\omega_1}{c}-n_e\left(\omega_2,\phi_0^{(II)}\right)\frac{\omega_2}{c},
\end{eqnarray}
\begin{eqnarray}
 \nonumber
 \Delta_{21}(\omega_1,\omega_2)=k_p\left(\omega_1+\omega_2, \phi_0^{(II)}\right)-k_o(\omega_2)-k_e\left(\omega_1, \phi_0^{(II)}\right)\\
 \label{Del21}
 =n_e\left(\omega_1+\omega_2,\,\phi_0^{(II)}\right)\frac{\omega_1+\omega_2}{c}
 -n_o(\omega_2)\frac{\omega_2}{c}-n_e\left(\omega_1,\phi_0^{(II)}\right)\frac{\omega_1}{c}.
\end{eqnarray}
Mismatches $\Delta_{12}(\omega_1,\omega_2)$ and $\Delta_{21}(\omega_1,\omega_2)$ are equal to each other only at $\omega_1=\omega_2$. Otherwise they are different. In particular, in dependence on one of the frequencies (e.g., $\omega_1$) at a given value of another one ($\omega_2$) mismatches $\Delta_{12}$ and $\Delta_{21}$ turn zero at different values of $\omega_1$. This is illustrated by Figure \ref{Fig5} where, as usual, $\Delta_{12}$ and $\Delta_{21}$ are considered as functions of wavelengths rather than frequencies. \ref{Fig5}).
Appearance of two different values of $\lambda_1$ (or, equivalently, $\omega_1$) indicates a possibility of observing two different peaks in the coincidence spectra. But, actually, one can observe either single-peak or double-peak coincidence spectra depending on conditions of observation.

Indeed, the usual way of observing coincidence spectra consists in splitting the SPDC biphoton beam by a nonselective beam splitter for two channels $A$ and $B$ with identical beams in each channel. Detectors (photon counters) are installed in each channel, and signals from each detector are sent to the computer to select only coinciding events, i.e., counts of photons of the same rather than different SPDC pairs. Parameters of one detector (e.g., in the channel $B$) are kept constant whereas parameters of the other detector ($A$) are scanned to reconstruct is such way the coincidence distribution of photons (number of counts) over the varying parameters. For finding spectral distributions detectors must be spectrally sensitive, i.e., having frequency filters installed in front of photon counters. If in addition detectors are provided with polarizers in front of photon counters, detectors are polarization-selective. In this case one can see separately only one of two coincidence peaks determined by the wave function (\ref{symmWF}). For example, if the detector in the channel $B$ counts only vertically polarized photons, and we denote their frequency as $\omega_2$, these photons belong to the type of extraordinary waves and are described by the first term in the sum of Eq. (\ref{symmWF}). Under these conditions the detector in the channel $A$ will count only horizomtally polarized (ordinary) photons, and its scanning over frequency $\omega_1$ (or wavelength $\lambda_1$) will reproduce the single coincidence spectral peak around the frequency (wavelength) where $\Delta_{12}=0$. Oppositely, if the detector in the channel $B$ is provided with the horizontal polarizer, the second detector (in the channel $A$) will register only vertically polarized photons, to reproduce the only coincidence peak described by the second term in Eq. (\ref{symmWF}) and located near the frequency (wavelength) where $\Delta_{21}=0$. To see both coincidence peaks together, one has to remove polarizers and to make detectors frequency-selective but non-sensitive to polarizations of photons.

The last case of measurements by only frequency-selective detectors corresponds in theory to construction of the density matrix from the wave function (\ref{symmWF}) and reduction (tracing) of this density matrix over the polarization variables. The arising frequency density matrix is given by
\begin{eqnarray}
 \nonumber
 \rho(\omega_1,\omega_2;\omega_1^\prime,\omega_2^\prime)\propto \exp\left\{-\left[(\omega_1+\omega_2-\omega_0)^2+(\omega_1^\prime+\omega_2^\prime-\omega_0)^2\right]\frac{\tau^2}{8\ln 2}\right\}\\
 \nonumber
 \times\Bigg\{{\rm sinc}\left[\frac{L}{2}\Delta_{12}(\omega_1,\omega_2)\right]{\rm sinc}\left[\frac{L}{2}\Delta_{12}(\omega_1^\prime,\omega_2^\prime)\right]\\
 \label{rho}
 +{\rm sinc}\left[\frac{L}{2}\Delta_{21}(\omega_1,\omega_2)\right]{\rm sinc}\left[\frac{L}{2}\Delta_{21}(\omega_1^\prime,\omega_2^\prime)\right]\Bigg\}.
\end{eqnarray}
This density matrix describes a mixed rather than pure state with continuous variables. Even definition of the degree of entanglement for such states is a problem having no clear and recognized solution. But measurable distributions of probabilities are rather easily found because they are directly determined by diagonal elements of the density matrix (\ref{rho}). The two-frequency probability density is given by
\begin{eqnarray}
 \nonumber
 \frac{dw}{d\omega_1d\omega_2}=\rho(\omega_1,\omega_2;\omega_1,\omega_2)\propto \exp\left\{-(\omega_1+\omega_2-\omega_0)^2\frac{\tau^2}{4\ln 2}\right\}\\
 \label{two-fr- probability}
 \times\Bigg\{{\rm sinc}^2\left[\frac{L}{2}\Delta_{12}(\omega_1,\omega_2)\right]+
 {\rm sinc}^2\left[\frac{L}{2}\Delta_{21}(\omega_1,\omega_2)\right]\Bigg\}.
\end{eqnarray}
Conditional probability density for one photon to have a frequencies $\omega_1$ under the condition that the other photon has some given frequency $\omega_2$, $dw^{(c)}/d\omega_1|_{\omega=const.}$, is determined by the same equation (\ref{two-fr- probability}). Alternatively, this equation can be rewritten in terms of wavelengths:
\begin{eqnarray}
 \nonumber
 \frac{dw^{(c)}}{d\lambda_1}=\exp\left\{-\left(\frac{1}{\lambda_1}+\frac{1}{\lambda_2}
 -\frac{1}{\lambda_0}\right)^2\frac{\tau^2}{4\ln 2}\right\}\\
 \label{probabilities-wavelengths}
 \times\Bigg\{{\rm sinc}^2\left[\frac{L}{2}\,\Delta_{12}\left(\frac{2\pi c}{\lambda_1},\frac{2\pi c}{\lambda_2}\right)\right]+
 {\rm sinc}^2\left[\frac{L}{2}\,\Delta_{21}\left(\frac{2\pi c}{\lambda_1},\frac{2\pi c}{\lambda_2}\right)\right]\Bigg\}.
\end{eqnarray}

\begin{figure}[h]
\centering\includegraphics[width=12.5cm]{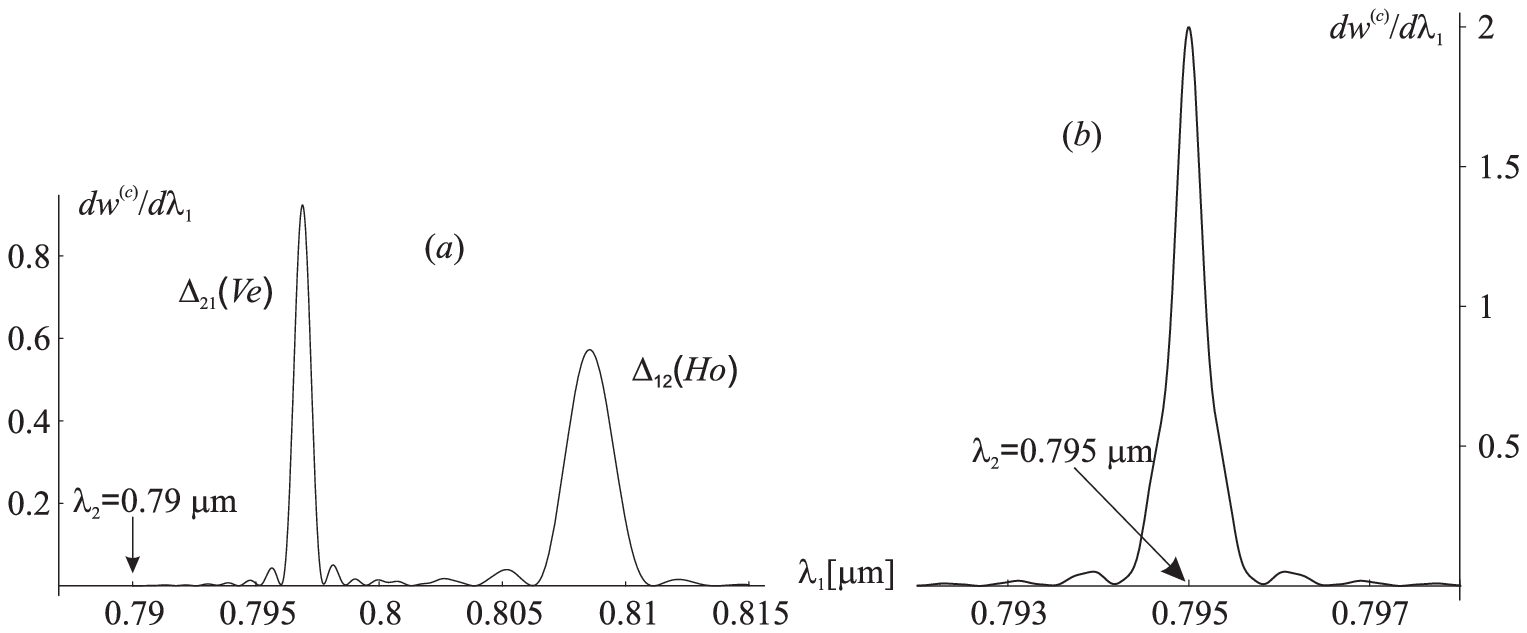}
\caption{{\protect\footnotesize {Coincidence spectra, $dw^{(c)}/d\lambda_1$ (in arbitrary units) at $\lambda_2=const. = (a)\, 0.79 \mu{\rm m}$ and $(b)\,0.795 \mu{\rm m}$; the type-II collinear degenerate phase-matching, pulse duration and central wavelength of the pummp $\tau=50 fs$ and $\lambda_0=0.795 \mu{\rm m}$, LiIO$_3$ crystal of the length $L=1 cm$.}}}\label{Fig6}
\end{figure}

\begin{figure}[h]
\centering\includegraphics[width=9cm]{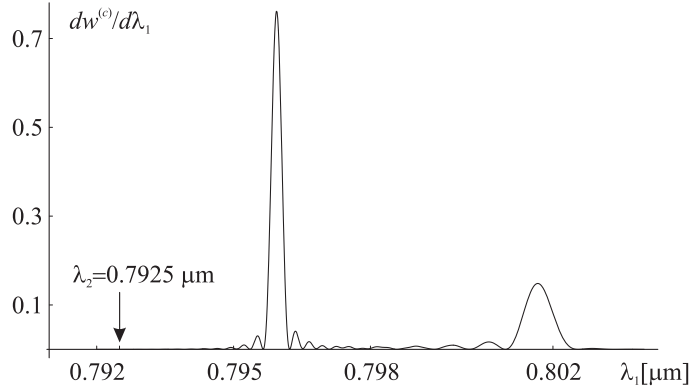}
\caption{{\protect\footnotesize {The same as in Figure \ref{Fig7}$(a)$, but at $\tau=186\,{\rm fs}$ and $\lambda_2=0.7925\,\mu{\rm m}$.}}}\label{Fig7}
\end{figure}

Of course, the theoretical concept of the conditional probability density is identical to the experimentally measurable coincidence distribution. Two typical examples of the double-peak coincidence spectra are shown in Figures \ref{Fig6}$a$ and \ref{Fig7}. These two pictures differ from each other by different pulse durations of the pump. As seen, in the case of longer pulses (Figure \ref{Fig7}) the double-peak structure is spectrally compressed compared to the case of longer pulses (Figure \ref{Fig6}$a$). The picture of Figure \ref{Fig6}$b$ shows how two peaks of the double-peak structure merge into a single peak when wavelengths (frequencies) of two emitted photons become equal to each other.


\section{Heights and widths of peaks and temporal walk-off}

Practically always both coincidence and single-particle SPDC spectra are narrow compared to the pump frequency $\omega_0$. For this reason wave vectors in the expressions (\ref{Del12}) and (\ref{Del21}) for phase mismatches can be expanded in powers of deviations of frequencies $\omega_1$ and $\omega_2$ from their central meanings $\nu_1=\omega_1-\omega_0/2$ and $\nu_2=\omega_2-\omega_0/2$:
\begin{eqnarray}
 \nonumber
 k_p(\omega_1+\omega_2)\approx k_p(\omega_0)+(\nu_1+\nu_2)k_p^\prime=k_p(\omega_0)+\frac{\nu_1+\nu_2}{{\rm v}_g^{(p)}}\\
 \nonumber
 k_e(\omega_{1,2})\approx k_e(\omega_0/2)+\nu_{1,2}k_e^\prime=k_e(\omega_0/2)+\frac{\nu_{1,2}}{{\rm v}_g^{(e)}},\\
 \label{expansion}
 k_o(\omega_{1,2})\approx k_o(\omega_0/2)+\nu_{1,2}k_o^\prime=k_e(\omega_0/2)+\frac{\nu_{1,2}}{{\rm v}_g^{(o)}},
\end{eqnarray}
where ${\rm v}_g^{(p,\,o,\,e)}$ are the group velocities of the pump and emitted ordinary and extraordinary photons
\begin{equation}
 \label{gr-vel}
 {\rm v}_g^{(p)}=\left(\left.\frac{dk_p(\omega)}{d\omega}\right|_{\omega=\omega_0}\right)^{-1},\;  {\rm v}_g^{(o,\,e)}\left(\left.\frac{dk_{o,\,e}(\omega)}{d\omega}\right|_{\omega=\omega_0/2}\right)^{-1}.
\end{equation}
In terms of $\nu_{1,2}$ the wave function of Eq. (\ref{symmWF}) takes the form
\begin{eqnarray}
 \nonumber
 \Psi\propto \exp\left(-\frac{(\nu_1+\nu_2)^2\tau^2}{8\ln 2}\right)\Bigg\{\left(1\atop 0\right)_1\left(0\atop 1\right)_2{\rm sinc}\left[\frac{L}{2c}
 (A_o\nu_1+A_e\nu_2)\right]\\
 \label{WF-nu}
  +\left(1\atop 0\right)_2\left(0\atop 1\right)_1{\rm sinc}\left[\frac{L}{2c}(A_o\nu_2+A_e\nu_1)\right]\Bigg\},
\end{eqnarray}
where $A_o$ and $A_e$ are the temporal-walk-off constants
\begin{equation}
 \label{walk-off-const}
 A_o=c\left(\frac{1}{{\rm v}_g^{(p)}}-\frac{1}{{\rm v}_g^{(o)}}\right),\;
 A_e=c\left(\frac{1}{{\rm v}_g^{(p)}}-\frac{1}{{\rm v}_g^{(e)}}\right).
\end{equation}
For LiIO$_3$ and $\lambda_0=0.3975 \mu{\rm m}$ the group velocities ${\rm v}_g^{(p,\,o,\,e)}$ and the walk-off constants $A_{o,\,e}$ are easily calculated to be given by: $v_g^{(p)}=0.4986\,c$, $v_g^{(o)}=0.522\,c$, $v_g^{(e)}=0.5628\,c$, $A_o=0.09$, and $A_e=0.2287$.

Let us assume now that that the pulse duration of the pump is so short that the Gaussian function in Eq. (\ref{WF-nu}) is much wider than both sinc-functions in their dependence on $\nu_1$, i.e., that
\begin{equation}
 \label{short pulses}
 \tau\ll \frac{LA_{o,\,e}}{c}\approx 3 ps.
\end{equation}
Then, positions and widths of the coincidence spectral peaks are determined mainly by the narrow sinc-functions. In dependence on $\nu_1$ at a given $\nu_2$, peaks related to mismatches $\Delta_{12}$ and $\Delta_{21}$ are located at $\nu_1^{(12)}$ and $\nu_1^{(21)}$ and have widths $\Delta\nu_1^{(12)}$ and $\Delta\nu_1^{(21)}$ given by
\begin{eqnarray}
 \label{positions-nu}
 \nu_1^{(12)} =-\nu_2\frac{A_e}{A_o}=-2.54\nu_2,\,
 \nu_1^{(21)}=-\nu_2\frac{A_o}{A_e}=-0.39\nu_2,\\
 \label{widths-nu}
 \Delta\nu_1^{(12)}\sim\frac{c}{LA_o},\,
 \Delta\nu_1^{(21)}\sim\frac{c}{LA_e}.
\end{eqnarray}
In terms of wavelengths the same relations take the form
\begin{eqnarray}
 \nonumber
 \lambda_1^{(12)}-2\lambda_0 =-(\lambda_2-2\lambda_0)\frac{A_e}{A_o}=-2.54(\lambda_2-2\lambda_0),\\
 \label{positions-lambda}
 \lambda_1^{(21)}-2\lambda_0=-(\lambda_2-2\lambda_0)\frac{A_o}{A_e}=-0.39(\lambda_2-2\lambda_0),\\
 \label{widths-lambda}
 \Delta\lambda_1^{(12)}\sim\frac{\lambda_0^2}{LA_o},\,
 \Delta\lambda_1^{(21)}\sim\frac{\lambda_0^2}{LA_e}.
\end{eqnarray}
These equations show that the peaks of the coincidence spectrum $dw/d\omega_1$ are located always at the opposite side of the central frequency $\omega_0/2$ or wavelength $2\lambda_0$  compared to the frequency $\omega_2$ (wavelength $\lambda_2$) of photons registered by the detector with fixed parameters (in the channel $B$). The peak located closer to the central frequency (wavelength) is described by the term with the mismatch $\Delta_{21}$ in Eq. (\ref{symmWF}) and is significantly narrower than the second wider peak, located almost ten times further from the central frequency (wavelength) and related to the term with $\Delta_{12}$ in Eq. (\ref{symmWF}). Both the double-peak structure and its described features arise owing to the well pronounced difference between the walk-off constants $A_o$ and $A_e$, and this difference originates from the fact that two emitted photons belong to different types of waves, ordinary ad extraordinary, with different group velocities. If the constants $A_o$ and $A_e$ would be equal, two peaks in the coincidence spectrum would merge into one, and this is the situation occurring in the type-I SPDC process, in which both emitted photons are ordinary waves.

As for heights of the peaks in the coincidence spectrum, as long as we do not calculate normalization factors in wave functions and density matrices, absolute values of the peak heights cannot be found. But in arbitrary units, in the approximation of short pump pulses (\ref{short pulses}), the peak heights are determined by the Gausssian function in Eqs. (\ref{symmWF}),(\ref{rho}),(\ref{two-fr- probability}),(\ref{WF-nu}) in which $\omega_1$ is substituted either by $\omega_0/2+\nu_1^{(12)}$ or by $\omega_0/2+\nu_1^{(21)}$:
\begin{equation}
 \label{peak-12}
 \left(\frac{dw}{d\omega_1}\right)_{\max}^{(12)}= \exp\left\{-\left(\nu_1^{(12)}+\nu_2\right)^2\frac{\tau^2}{4\ln 2}\right\}
 =\exp\left\{-\frac{(A_e-A_o)^2}{A_o^2}\frac{\nu_2^2\tau^2}{4\ln 2}\right\},
\end{equation}
\begin{equation}
 \label{peak-21}
 \left(\frac{dw}{d\omega_1}\right)_{\max}^{(21)}= \exp\left\{-\left(\nu_1^{(21)}+\nu_2\right)^2\frac{\tau^2}{4\ln 2}\right\}
 =\exp\left\{-\frac{(A_e-A_o)^2}{A_e^2}\frac{\nu_2^2\tau^2}{4\ln 2}\right\}.
\end{equation}
The ratio of peak heights equals to
\begin{equation}
 \label{peak-ratio}
 \frac{(dw/d\omega_1)_{\max}^{(21)}}{(dw/d\omega_1)_{\max}^{(12)}}
 =\exp\left\{\frac{(A_e^2-A_o^2)(A_e-A_o)^2}{A_o^2A_e^2}\frac{\nu_2^2\tau^2}{4\ln 2}\right\}.
\end{equation}
If this ratio is large, the farthest peak becomes practically invisible. On the other hand, the ratio (\ref{peak-ratio}) can be minimized by making the fixed frequency $\nu_2$ sufficiently small. The question that arises in such case is how small is a reasonable value of $|\nu_2|$? The answer is evident: the minimal value of $|\nu_2|$ is that at which the distance between peaks becomes on the order of width of the farthest peak, i.e., $|\nu_2|_{\min}\sim\frac{c}{L}\frac{A_eA_o}{A_e^2-A_o^2}$. Substitution of this value into the right-hand side of Eq. (\ref{peak-ratio}) gives the minimal achievable value of the peak-height ratio:
\begin{equation}
 \label{peak-ratio-min}
 \left[\frac{(dw/d\omega_1)_{\max}^{(21)}}{(dw/d\omega_1)_{\max}^{(12)}}\right]_{\min}
 =\exp\left\{\frac{(A_e-A_o)}{A_o^2(A_e+A_o)}\frac{c^2\tau^2}{4L^2\ln 2}\right\}.
\end{equation}
Under the condition (\ref{short pulses}) the expression under the symbol of exponent in this equation is small and, hence, the peak-height ratio is close to unit. This means that, indeed, in the femtosecond region of pulse durations of the pump the double-peak structure of the coincidence spectrum is rather well
pronounced in a sufficiently large region of fixed frequencies  $|\nu_2|$ exceeding $|\nu_2|_{\min}$ and is accessible for experimental observation.

\section{Conclusion}

Thus, the effect described above consists in appearance of a double-peak structure in the coincidence spectrum of the biphoton state arising in the SPDC process with a collinear type-II phase-matching, short duration of pump pulses and degeneracy with respect to central frequencies of emitted photons. The effect is shown to be related directly to the symmetry of the biphoton wave function with respect to the transposition of photon variables. Though often not noticed or ignored, this feature of two-bozon wave functions is shown to be responsible for a clearly observable and not negligible effect of peak doubling.  The conditions under which the effect can be observed are determined. The main of them consists in the requirement that the pump pulses have to be short enough, of a femtosecond duration. Positions, heights and widths of the peaks are found and expressed in terms of the temporal-walk-off parameters of the SPDC process.

\section{Acknowledgement}
The work is supported partially by the RFBR grants 08-02-01404, 10-02-90036, and 11-02-01043

\end{document}